\documentclass[10pt]{article}

\def\block{{\rm block}}
\def\blocks{{\rm blocks}}
\begin{document}

\noindent{\Large \bf
The Renormalization-Group peculiarities of Griffiths and Pearce:
\break
What have we learned?
\normalsize \rm}

\vspace{24pt}

\noindent{\bf  Aernout C.D. van ENTER$^{1}$, 

\vspace{12pt}
\it
\noindent
$^{(1)}$ Instituut voor theoretische natuurkunde R.U.G.
Nijenborgh 4, 9747AG, Groningen, the Netherlands.
\rm

\vspace{24pt}
\footnotesize
\begin{quote}
{\bf Abstract:} 



We review what we have learned about the ``Renormalization Group 
peculiarities'' which were discovered about twenty years ago 
by Griffiths and Pearce, and which questions they asked are still widely open.
We also mention some related developments.

\vspace{3pt}


\end{quote}
\normalsize

\vspace{12pt}

\section{Introduction}

About twenty years ago, Griffiths and Pearce \cite{gripea78,gripea79} 
discovered some unexpected
mathematical difficulties in rigorously implementing many of the generally
used real-space Renormalization Group transformations. 

In this contribution I plan to 
assess what we have learned about these problems since then.
  In particular  we will see that, although it turns out that Renormalization-Group maps
cannot be discontinuous, they can be ill-defined. This means that in such 
cases no ``reasonable'' renormalized Hamiltonian can be found.

Moreover, in any region 
of the phase diagram the question whether a particular transformation is 
well-defined or ill-defined turns out to be highly non-trivial. 
The ill-definedness
of various Renormalization-Group maps can be expressed in the violation 
of the property of ``quasilocality'' in the renormalized states. The 
study and classification of such non-quasilocal states (---non-Gibbsian 
measures---)
has led also to various results of mathematical interest, some of which we will
mention further on.  
Some  other papers covering the area of non-Gibbsianness and 
Renormalization-Group peculiarities, treating
further related material can be found in 
\cite{vEFS94,vEFS_JSP,vEbud,lordis95,veldis95,fer}
and references therein.

\section{Gibbs measures and quasilocality}

In this section we will describe some definitions and facts we will need about 
the theory of Gibbs measures. For a more extensive treatment we refer to 
\cite{geo88} or \cite{vEFS_JSP}.

We will consider spin systems on a lattice $Z^d$, where in most cases we will
take a single-spin space $\Omega_{0}$ which is finite. The configuration space 
of the whole system is $\Omega =\Omega_{0}^{Z^d}$. Configurations will be 
denoted by $\sigma$ or $\omega$, and their coordinates at lattice site i are
denoted by $\omega(i)$ or $\sigma(i)$. A (regular) interaction $\Phi$ is a 
collection of functions $\Phi_{X}$ on $\Omega_{0}^X, X \in Z^d$ which is
translation invariant and satisfy:
\begin{equation}
\Sigma_{0 \in X} \  |\Phi_{X}|_{\infty} < \infty
\end{equation}
Formally Hamiltonians are given by
\begin{equation}
\ H^{\Phi} = \Sigma_{X \in Z^d} \  \Phi_{X}
\end{equation}
Under the above regularity condition these type of expressions make 
mathematical sense if the
sum is taken over all subsets having non-empty intersections with a finite
volume $\Lambda$. For regular interactions one can define Gibbs measures as
probability measures on $\Omega$ having conditional probabilities which are
described in terms of appropriate Boltzmann-Gibbs factors:
\begin{equation}
{\mu(\sigma_{\Lambda}^{1}|\omega_{\Lambda^{c}})\over{{\mu(\sigma_{\Lambda}^{2}|\omega_{\Lambda^{c}})}}} = 
\exp - (\Sigma_{X} \ [\Phi_{X}(\sigma_{\Lambda}^{1} \omega_{\Lambda^{c}}) \ - 
\ \Phi_{X}(\sigma_{\Lambda}^{2} \omega_{\Lambda^{c}})])
\end{equation}
for each volume $\Lambda$, $\mu$-almost every boundary condition 
$\omega_{\Lambda^{c}}$ outside $\Lambda$ and each pair of configurations
$\sigma_{\Lambda}^{1}$ and $\sigma_{\Lambda}^{2}$ in $\Lambda$.
As long as $\Omega_{0}$ is compact, there always exists at least one
Gibbs measure for every regular interaction; the existence of more than one
Gibbs measure is one definition of the occurrence of a first-order phase
transition of some sort. Every Gibbs measure has the property that (for one of 
its versions) its conditional probabilities are continuous functions of the 
boundary condition $\omega_{\Lambda^{c}}$. It is a non-trivial fact that this 
continuity, which goes by the name ``quasilocality'' or ``almost 
Markovianness'', in fact characterizes the Gibbs measures \cite{sul,koz}, once 
one knows that all the conditional probabilities are bounded away from zero
(that is, the measure is {\em nonnull} or has the {\em finite energy} 
property). In some examples it turns out to be possible to check this
continuity (quasilocality) property quite explicitly.  
If a measure is a Gibbs measure for a regular interaction, this interaction
is essentially uniquely determined.


A second characterization of Gibbs measures uses the variational principle
expressing that in equilibrium a system minimizes its free energy. A 
probabilistic formulation of this fact naturally occurs in terms of the theory 
of large deviations. A (third level) large deviation rate function is up to a 
constant and a sign equal to a free energy density. 
To be precise, let $\mu $ be a translation invariant Gibbs measure, and 
let $\nu $ be an arbitrary translation invariant measure.
Then the relative entropy density $i( \nu |\mu )$ can be defined as the limit:
\begin{equation}
i(\nu |\mu ) = lim_{\Lambda \rightarrow Z^d} \ {1 \over |\Lambda |} \ I_{\Lambda}(\nu |\mu)
\end{equation}
where
\begin{equation}
I_{\Lambda}(\nu |\mu)= \int log ({d\nu_{\Lambda} \over d\mu_{\Lambda}}) d\nu_{\Lambda}  
\end{equation}
and $\mu_\Lambda$ and $\nu_\Lambda$ are the restrictions of $\mu$ and $\nu$
to $\Omega_0^\Lambda$.
It has the property that $i( \nu |\mu )= 0$ if and only if the measure $\nu$
is a Gibbs measure for the same interaction as the base measure $\mu$.
We can use this result in applications if we know for example that a known 
measure $\nu$ cannot be a Gibbs 
measure for the same interaction as some measure $\mu$ we want to investigate. 
For example, if $\nu$ is a 
point measure, or if it is the case that $\nu$ is a product measure and $\mu$ 
is not, we can conclude from the statement:  $i( \nu |\mu )= 0$, that $\mu$
lacks the Gibbs property.

For another method of proving that a measure is non-Gibbsian because of having 
the ``wrong'' type of (in this case too small) large deviation probabilities, 
see \cite{sch87}.   

\section{Renormalization-Group maps: some examples}
We will mostly consider the standard nearest neighbor Ising model with (formal)
Hamiltonian
\begin{equation}
\ H = \Sigma_{<i,j>} \;  -\sigma(i) \sigma(j) \; -h \Sigma_{i}\; \sigma(i)
\end{equation}
at inverse temperature $\beta$. The magnetic field strength is $h$. 
The dimension $d$ in what follows will be at least 2.

We will consider various real-space Renormalization-Group or block-spin 
transformations which 
act on the Ising Gibbs measures. The question then is to find the renormalized 
interaction, that is the interaction associated to the transformed measure.

Although in applications the transformation 
needs to be iterated we will mostly restrict ourselves to considering
a single transformation. Existence of the first step is of course necessary
but far from sufficient for justification of an iterative procedure.
As a further remark we mention that sometimes even if the first step is ill-defined,
after repeated transformations a transformed interaction can be found\cite{maroli93}.

We divide the lattice into a collection of
non-overlapping blocks.
A Renormalization-Group transformation defined at the level of measures 
will be a probability kernel
\begin{equation}
T(\omega'| \omega)= \Pi_{\blocks}T(\omega'(j)|\omega(i);\ i \  \in \block_j)
\end{equation} 
This means that the distribution of a block-spin  
depends only on the spins in the corresponding block, in other words the transformation 
is local in real space. The case of a deterministic transformation is included,
by having a T which is either zero or one.

Renormalization-Group methods are widely in use to study phase transitions
and in particular critical phenomena of various sorts (see for example 
\cite {kogwil,ma,domgre,fis98}). Some good more recent references in which the theory
is explained, mostly at a physical level of rigour, but with more careful statements
about what actually has and has not been proven are  \cite{gol,bengal}.

\bigskip

1) One class of examples we will consider are (linear) block-average 
transformations. This means that the block-spins are the average spins 
in each block. 
Applied to Ising systems they suffer from the
fact that the renormalized system has a different single-spin space from
the original one. Despite this objection, the linearity makes these maps
mathematically rather attractive, and they have often been considered.
Because we are not iterating the transformation we need not worry too much
about the single-spin space changing.

\bigskip

2) Majority rule and Kadanoff transformations.

In the case of majority rule transformations \cite{nievle} applied to  
blocks containg an odd number of sites, the block spin is just given by 
the sign of the majority 
of the spins in the block. These transformations have been chosen often
because of their numerical tractability.

The Kadanoff transformation is a soft version (a proper example of a 
stochastic transformation) of the majority rule:
\begin{equation} 
T(\sigma'(j)|\sigma(i); \ i \in \block_{j}))={C \  \exp \ [p \sigma'(j) \Sigma_{i \in 
\block_{j}} \ \sigma(i)]}  
\end{equation}
In the limit in which $p$ goes to $\infty$ the Kadanoff map reduces to 
a majority rule transformation.

\bigskip

3)Decimations and projections.

We will call a ``decimation'' taking the marginal of a measure restricted to 
the spins on a sublattice of the same dimension as the original system, 
(thus the block-spins are the spins in some periodic sub-lattice).

(A ``projection'' will mean taking the marginal to a lower-dimensional 
sublattice. Projections are not Renormalization-Group maps proper, but share 
some mathematical properties of Renormalizatio-Group maps. See \cite{mamore,mamore1})

Although  decimation transformation have the advantages both of being  linear 
and of preserving the 
single-spin space, 
infinite iteration has the disadvantage that critical fixed points won't
occur. However, this problem does not show up after a finite number of 
applications of these maps, so we will not need to worry about it.

\section{The investigations of Griffiths and Pearce}
 
Griffiths and Pearce \cite{gripea78, gripea79} seem to be the first 
investigators who looked seriously at the question 
whether renormalized Hamiltonians exist in a precise mathematical sense.
They found that for some real-space transformations like decimation or 
Kadanoff transformations 
in the low-density regime (that means strong magnetic fields in Ising language) the 
Renormalization-Group map maps the Ising Gibbs measure on a Gibbs measure for 
an in principle computable interaction. 
They also found, both at phase transitions and near areas of the phase 
diagram where first order transitions occur, that there are regimes where
the formal expression for renormalized interactions behaved in a peculiar way.
These ``peculiarities'' were found for decimation, Kadanoff, and majority rule
transformations. The problem underlying the peculiarities is the occurrence of phase
transitions in the system, once it is constrained by prescribing some 
particular, rather 
atypical, block-spin configuration. This means that there can exist for these
block-spin configurations long-range correlations in the presumed 
``short-wavelength degrees of freedom''- or ``internal spins''-,
which are to be integrated out in a Renormalization-Group map.

In their paper Griffiths and Pearce discuss various possible explanations of
these ``peculiarities'':
\smallskip

\noindent
P1)The renormalized interaction might not exist, 

\smallskip

\noindent
P2)it might exist but be a singular function of the original interaction, 

\smallskip

\noindent
P3)it might be non-quasilocal, 

\noindent
or 

\smallskip

\noindent
P4)the thermodynamic limit might be problematic.

Shortly after, the problem was studied by Israel \cite{isr79}. He
obtained (very)high temperature existence results, including approach to 
trivial fixed points, as well as an analysis
of the decimation transformation at low temperature, indicating strongly
that in that case the renormalized interaction does not exist. 
Israel's results convinced Griffiths that in fact possibility P1)--- 
non-existence
of the renormalized interaction--- applies \cite{gri81}, however, it seems that
most authors aware of their work interpreted the Griffiths-Pearce
peculiarities as singular behaviour of the renormalized interactions
(possibility P2)). See for example \cite{hashas, burlee,hud, thom}.
Many authors  did not even show awareness that the dtermination of renormalized 
interactions presented problems beyond mere computational difficulty.
See the Appendix for some illustrations of this point.

Deep in the uniqueness regime the Renormalization-Group procedure appeared
to be well-defined in some generality.

At that stage, various open questions formulated by Griffiths and Pearce and
Israel were still left:

\smallskip

\noindent
Q1) What is the nature of the ``peculiarities''?

\smallskip

\noindent
Q2) Can one say anything about the critical regime?

\smallskip

\noindent
Q3) Do different transformations exhibit similar behaviour?
In particular are momentum-space Renormalization-Group transformations similar to real-space
transformations  regarding the occurrence of ``peculiarities''?

\smallskip

\noindent
Q4) As the ``peculiarities'' are due to rather atypical spin-configurations,
can one make the Renormalization-Group enterprise work, by considering
only typical configurations, and thus work with appropriate approximations? 
Or, more generally, is there a framework in which one can one 
implement the whole Renormalization-Group 
machinery in a mathematically correct way? 
\smallskip

\section{Answered and unanswered questions} 

About question Q1 ---the nature of the peculiarities--- we have acquired
some more insight.
In \cite{vEFS_JSP} the Griffiths-Pearce study was taken up and further
pursued. It was found, making use of the above-mentioned 
variational characterization 
of Gibbs measures, that the peculiarities could {\em not} be due to 
discontinuities
in the Renormalization-Group maps, and that in the cases considered by
Griffiths and Pearce the renormalized measures all have conditional
probabilities with points of essential discontinuity. That is, they are {\em
non-Gibbsian}. See also \cite{ken}. Thus a renormalized Hamiltonian 
{\em does not exist.} This despite many attempts to compute 
these ---non-existent--- renormalized Hamiltonians, and the various physically 
plausible and intuitively convincing conclusions, derived from such 
approximate computations. 

In fact, those constrained block-spin configurations, pointed out by
Griffiths and Pearce, for which the internal
spins have phase transitions 
are precisely the points of discontinuity -- non-quasilocality-- 
for some conditional probability. The observation that the ``peculiarities''
were due to the violation of the
quasilocality condition  was in essence, 
though somewhat in a slightly implicit way, already made in Israel's analysis.

At first the non-Gibbsian examples were found at or near phase transitions,
at sufficiently low temperatures. Then it was found  \cite{vEFK_JSP} that 
decimation
applied to many-state Potts models gives non-Gibbsian measures also 
above the transition temperature.

Somewhat surprisingly, it turned out that even deep in the  uniqueness regime,
the non-Gibbsianness can occur. This happens for example at low temperature for
block-average \cite{vEFS_JSP}, and majority rule \cite{vEFK_JSP} 
transformations in 
strong external fields, and it is even possible to devise transformations for
which this happens at arbitrarily high temperatures \cite{vE97}. 
On the other hand, it turns out that for Gibbs measures well in the uniqueness regime,
a repeated application of  a decimation 
transformation, even after composing with another 
Renormalization-Group map, leads again to a 
Gibbs measure, although applying the decimation only a few times may result
in a non-Gibbsian measure \cite{maroli93,maroli95}. For related work in this
direction 
see also \cite{BCO}. Physically, this means that one cannot find a 
``reasonable'' (more or less short-range, local) Hamiltonian. 

\smallskip

About critical points (question Q2), Haller and Kennedy \cite{halken95} 
obtained the first 
results, proving both for a decimation and a Kadanoff transformation  example 
that  a single Renormalization-Group map can map an area including a critical
point to  a set of renormalized interactions. There are strong indications 
for similar
behaviour for other transformations
\cite{ken92,benmaroli,ciroli96,O-L96}. The indications are partly numerical,however,
and fall short of a rigorous proof. See also the recent numerical work of 
\cite{Pinn}. On the other hand, counterexamples where a transformed critical
measure is non-Gibbsian also exist \cite{vE97,vEbud}

Another critical regime result is the observation of \cite{dorvE}, 
that majority rule scaling limits of critical points above the upper critical
dimension (when the critical behaviour is  like that of a Gaussian) are non-Gibbsian.

\smallskip
Recently it was found that non-Gibbsian measures can also occur as a result of 
applying momentum-space transformations \cite{vEF99}. 
The conclusion of all the above is
that different transformations can have very different behaviour.This is 
a sort of answer to question Q3, although in principle not a very 
informative one. In fact for applicability, if not for existence, something 
like this was already expected (compare for example Fisher's \cite{fis83}
remarks on ``aptness'' and focusability).

Regarding question Q4) ---to find the right setting for implementing 
Renormalization-Group Theory--- the issue is still essentially open. One approach
which was stimulated by the late R.L. Dobrushin is related
to the observation of Griffiths and Pearce that the block-spin configurations
responsible for the peculiarities (the discontinuity points) are rather
atypical. By removing them from configuration space, one might hope to
be left with a viable theory. Such investigations have led to the notions
of ``almost'' or ``weak'' Gibbs measures, the study of whose properties is 
being actively pursued \cite{dobsh1,dobsh2,brikule,mamore,mamore1, vES,MaSh,ferpfi97,
sep,lef98,maevel97,lorma}. This approach is somewhat along the lines of Griffiths' 
and Pearce's possibility P3). See also the next section.
Whether it is possible to describe Renormalization-Group
flows in spaces of  such interactions, while keeping a continuous connection 
between interactions describing a positively
and a negatively magnetized state, is not at all clear.

As for projections, it is known that on the phase-transition line of the 
2-dimensional Ising model the projection to $Z$ of any Gibbs measure is 
non-Gibbsian \cite{sch87}. In the whole uniqueness regime, except possibly at 
the critical point, this projection results in a Gibbsian measure 
\cite{maevel92,lordis95,lor95}. In three dimensions the projected measures 
to two-dimensional planes are 
again non-Gibbsian on the transition line \cite{ferpfi97,maevel94,mamore1}, however, 
now, due to the presumably existing surface (layering) transition between 
different Basuev states, one expects that the projected 
measures also in a small field will be non-Gibbsian \cite{lordis95,lor95}. 

The composition of a projection and a decimation in the phase transition region
gives rise to a new phenomenon, namely the possibility of a state-dependent
result. The transformed plus and minus measures are Gibbs measures for 
different interactions, while the transformed mixed measures are non-Gibbsian
\cite{lorvel94}.

\section{Further results on non-Gibbsian measures}

Further investigations in which   non-Gibbsian measures were displayed,
have been 
done about the random-cluster models of Fortuin and Kasteleyn 
\cite{pfivel95,gri95,borcha95,Hag5,sep}, about the Fuzzy Potts 
image analysis model 
\cite{maevel95}, and about various 
non-equilibrium models \cite{spe94,veldis95,marsco,maroli95}.

\bigskip




The non-Gibbsian character of the various measures considered comes
often as an unwelcome surprise. A description in terms of effective, 
coarse-grained or renormalized potentials is often convenient, and
even seems essential for some applications. Thus, the fact that such a 
description is
not available thus can be a severe drawback. As remarked earlier 
there have been attempts to
tame the non-Gibbsian pathologies, and here we want to give a short
comparison of how far one gets with some of those attempts.

\bigskip

 1) The fact that the constraints which act as points of discontinuity
often involve configurations which are very untypical for the measure under 
consideration, suggested a
notion of  {\em almost} Gibbsian or {\em weakly} Gibbsian measures. These are 
measures whose
conditional probabilities  are either continuous only on a set of full measure
or can be written in terms of an interaction which is summable only 
on a set of full measure. Intuitively, the difference is that 
in one case the 
``good'' configurations can shield off {\em all} influences from infinitely
far away, and in the other case only {\em almost all} influences.
The weakly Gibbsian 
approach was first suggested by Dobrushin to various people; his own version
was published only later\cite{dob95,dobsh1,dobsh2}. An early definition of
almost Gibbsianness 
appeared in print in \cite{lorwin92},see also \cite{ferpfi97},  \cite{maevel95, maevel97,mamore} and 
\cite{pfivel95} for further developments. Some examples of measures which are at the worst almost 
Gibbsian measures in this sense are decimated or projected Gibbs
measures in an external field, random-cluster measures on regular lattices, 
and low temperature fuzzy Potts measures. In the random-cluster measures one
can actually identify explicitly all bond configurations which give rise to 
the non-quasilocality. They are precisely those configurations in which 
(possibly after a local change) more than one infinite cluster coexist.

On a tree, because of the possible coexistence of infinitely many infinite 
clusters with positive probability, the random-cluster measure can violate the 
weak non-Gibbsianness condition and be strongly non-Gibbsian \cite{Hag5}. 

Dobrushin \cite{dob95,maevel97,dobsh1,dobsh2}
showed that for a projected pure phase on the coexistence line of the 
2-dimensional Ising model it is possible to find an almost everywhere defined 
interaction, hence these measures are weakly Gibbsian. 
His approach, which is via low-temperature expansions, provides 
a way of obtaining good control for the non-Gibbsian projection. 
For similar ideas in a Renormalization-Group setting see \cite{brikule,
lef98,lormaevel98}.

For some
Renormalization-Group examples an investigation via low temperature 
expansions into
the possibility of recognizing non-Gibbsianness was started in \cite{sal95}.

Another  simple counter-example of a strongly non-quasilocal measure, where 
each configuration can act as a point of
discontinuity, is a mixture of two Gibbs measures for different interactions
\cite{loren95}. 

\bigskip

2) Stability under decimation (and other transformations).

In \cite{maroli93,maroli95} it was shown how decimating once renormalized
non-Gibbsian measures results in Gibbs measures again after a sufficiently
large number of iterations. These often decimated measures are in the 
high-temperature regime, in which the usually applied Renormalization-Group
maps are well-defined (this does not hold true for all maps though, in view of 
the highly non-linear example given before). 

On the other hand, in examples
where the non-Gibbsian property is associated with large deviation properties
which are not compatible with a Gibbsian character (this holds for example
for projected Gaussians, invariant measures of the voter and the 
Martinelli-Scoppola model\cite{lebmae87,lebsch,marsco}) the non-Gibbsian property survives all sort of
transformations \cite{maroli95,loren95}. The argument is that when some
obviously non-Gibbsian measure has a rate function zero with respect to the 
measure under
consideration, this property is generally preserved (\cite{vEFS_JSP} 
Section 3.2 and 3.3).

The family of projected Gaussians include scaling
limits for majority-rule transformations in high dimensions \cite{dorvE}.
The transformation of those scaling limits is,heuristically,
interpreted as  making a move from a fixed 
point in what is usually called a ``redundant'' direction (cf. Wegner's 
contribution to \cite{domgre}) in some space of Hamiltonians. Here, of 
course, the whole point is that such Hamiltonians do not exist.    

\bigskip

3) The two criteria mentioned above are distinct. A simple one-dimensial example due to J. van den Berg 
\cite{lormaevel98} gives a
one-dependent measure which has a set of discontinuity points of full measure, 
but due to the one-dependence the measure becomes after decimation independent
and therefore trivially Gibbsian. In the opposite direction, there exist 
examples of measures whose non-Gibbsianness is robust, although they are weakly, and even almost Gibbsian\cite{vES}.

\section{Conclusions}    
We have by now managed well to understand the Griffiths-Pearce
peculiarities in the sense that we can identify mathematically their nature.
However, how to get around them, and make Renormalization-Group theory
mathematically respectable, is still a task requiring a lot of further work.
Renormalization-Group ideas have of course often been inspirational,
also for various rigorous analyses. Renormalization-Group implementations
on spin models, which both for numerical and pedagogical convenience
have often been
treated by Renormalization-Group methods, 
run into difficulties which seem hard to avoid. 
An implementation of Renormalization-Group ideas on contour models looks
more promising, at least for the description of first-order phase transitions
\cite{gkk,bovkul}.

\bigskip

In non-equilibrium statistical mechanics there are still many open questions
about the occurrence of non-Gibbsian measures. The term non-Gibbsian or 
non-reversible is often used for invariant measures in systems in which there 
is no detailed balance \cite{lig,eylebspo,ern95}. It is an open question
to what extent  
such measures are non-Gibbsian in the sense we have described here.
It has been conjectured that such measures for which there is no detailed 
balance are  quite generally non-Gibbsian in systems with a 
stochastic dynamics, see for example \cite{lebsch} or \cite{eylebspo}, 
Appendix 1; on the other hand it has been predicted
that non-Gibbsian measures are rather exceptional (\cite{lig}, Open problem 
IV.7.5, p.224) at least for non-reversible spin-flip processes under the 
assumptions of rates 
which are bounded away from zero. The examples we have are for the moment too 
few to develop a good intuition on this point.

Non-Gibbsian measures occur in quite different areas of statistical 
mechanics, and can have quite different properties. By now it seems somewhat 
surprising that it took so long to appreciatee the fact that the Gibbs property is 
rather special, in particular in view of Israel's \cite{isr92} result that in 
the set of all translation invariant (ergodic, nonnull) measures, Gibbs 
measures are exceptional. 

\section{Appendix:  Some literature}
As an illustration that the heuristic character of  Renormalization-Group 
theory, and in particular the need to consider the existence problem
of (approximately) local renormalized interactions was recognized by
various practitioners I mention the following quotes:

\noindent
One cannot write a renormalization cook book (K.G. Wilson, cited by 
Niemeijer and van Leeuwen \cite{domgre})

\noindent
The notion of renormalization group is not well-defined (\cite{bengal}, opening
sentence).

\noindent
It is dangerous to proceed without thinking about the physics \cite{gol}

\noindent
...the locality [of the renormalized interactions] is a non-trivial
problem which will not be discussed further \cite{kogwil}.

\noindent
A Renormalization Group for a space of Hamiltonians should satisfy the following \cite{fis83}:

\noindent
A) Existence in the thermodynamic limit,...

\noindent
C) Spatial locality...

\noindent
In the words of Lebowitz:
On the cautionary side one should remember that there are still some serious
open problems concerning the nature of the RG transformation of Hamiltonians
for statistical mechanical systems, i.e. for critical phenomena. A lot of 
mathematical work remains to be done to make it into a well-defined theory 
of phases transitions \cite{lebpr}.
\bigskip

As an illustration that on the other hand the occurrence of the 
above-mentioned problems has 
not been generally recognized, leading to some   incorrect or at 
least misleading statements, I quote:
\noindent
The renormalization-group operator ... transforms an arbitrary system in
the [interaction] space ... into another system in the space .... \cite{thom}

\noindent
[the set of coupling constants gives rise]..the most general form of the Hamiltonian...
\cite{hua}

\noindent
Further iterations of the renormalization group will generate long-range and 
multi-spin interactions of arbitrary complexity \cite{yeo}.

\noindent
...the space of Ising Hamiltonians in zero field may be specified by the set
of all possible spin-coupling constants, and is {\em closed} [Fisher's 
emphasis] under the decimation transformation \cite{fis98}.
\bigskip

\section*{Acknowledgments}
My contribution reflects mainly joint work.  I am grateful for all the 
collaborations I participated in, especially to Roberto
Fern{\'a}ndez and   Alan Sokal, and also to Tonny Dorlas, Roman Koteck{\'y},  
J{\'o}zsef L{\"o}rinczi, and Senya Shlosman, for all they 
taught me during these collaborations. 
Also I thank Roberto Fern\'andez and Marinus Winnink for helpful advice on the manuscript.
I have much benefitted from 
conversations and correspondence with many other colleagues. 
I thank the organisers for a very interesting meeting. 
This research has been supported by EU contract CHRX-CT93-0411,
and het Samenwerkingsverband Mathematische Fysica FOM{/}SWON.



\end{document}